\documentclass[times,twocolumn,final]{elsarticle}

\usepackage{cag}
\usepackage{framed,multirow}
\usepackage{hyperref}
\usepackage{booktabs}
\usepackage{amssymb}
\usepackage{latexsym}
\usepackage{lipsum}  
\usepackage{times}
\usepackage{epsfig}
\usepackage{graphicx}
\usepackage{amsmath}
\usepackage{float} 
\usepackage{subcaption}
\usepackage{enumitem}
\usepackage{array}
\usepackage{mathtools}
\usepackage{xspace} 
\usepackage[]{xcolor}
\usepackage{float} 
\usepackage{caption}

\definecolor{newcolor}{rgb}{.8,.349,.1}

\definecolor{amethyst}{rgb}{0.6, 0.4, 0.8}
\definecolor{darkpastelgreen}{rgb}{0.01, 0.75, 0.24}
\definecolor{amber}{rgb}{1.0, 0.75, 0.0}
\definecolor{cadmiumorange}{rgb}{0.93, 0.53, 0.18}
\definecolor{lawngreen}{rgb}{0.49, 0.99, 0.0}
\definecolor{limegreen}{rgb}{0.2, 0.8, 0.2}
\definecolor{neongreen}{rgb}{0.22, 0.88, 0.08}
\definecolor{amethyst}{rgb}{0.6, 0.4, 0.8}
\definecolor{darkpastelgreen}{rgb}{0.01, 0.75, 0.24}
\definecolor{greenbest}{RGB}{88,137,15}
\definecolor{redworst}{RGB}{137,15,27}
\definecolor{redpaper}{RGB}{196,77,88}
\definecolor{greenpaper}{RGB}{88,137,15}

\newcommand{\GreenColor}[1]{\textcolor{greenpaper}{\textbf{#1}}}
\newcommand{\RedColor}[1]{\textcolor{redpaper}{\textbf{#1}}}

\newcommand{\REMOVE}[1]{{}}

\newcommand{\Loss}{\mathcal{L}}

\newcommand{\bsdf}{\textsc{bsdf}}
\newcommand{\FLIP}{\protect\reflectbox{F}LIP\xspace}

\definecolor{newcolor}{rgb}{.8,.349,.1}

\journal{Computers \& Graphics}

\begin{document}

\verso{Preprint}

\begin{frontmatter}

\title{Single-image Reflectance and Transmittance Estimation from Any Flatbed Scanner}

\author[1,2,3]{Carlos \snm{Rodriguez-Pardo}\corref{cor1}}
\cortext[cor1]{Corresponding author. Work done at Universidad Rey Juan Carlos. }
\emailauthor{carlos.rodriguezpardo.jimenez@gmail.com}{Carlos Rodriguez-Pardo}
\author[4]{David \snm{Pascual-Hernandez}}
\author[5]{Javier \snm{Rodriguez-Vazquez}}
\author[4]{Jorge \snm{Lopez-Moreno}}
\author[4,6]{Elena \snm{Garces}}

\address[1]{Politecnico di Milano, Department of Management, Economics and Industrial Engineering}
\address[2]{Euro-Mediterranean Center on Climate Change (CMCC)}
\address[3]{RFF-CMCC European Institute on Economics and the Environment (EIEE)}
\address[4]{Universidad Rey Juan Carlos}
\address[5]{Arquimea Research Center}
\address[6]{Adobe Research}

\received{\today}

\begin{abstract}
Flatbed scanners have emerged as promising devices for high-resolution, single-image material capture. However, existing approaches assume very specific conditions, such as uniform diffuse illumination, which are only available in certain high-end devices, hindering their scalability and cost. In contrast, in this work, we introduce a method inspired by intrinsic image decomposition, which accurately removes both shading and specularity, effectively allowing captures with any flatbed scanner. Further, we extend previous work on single-image material reflectance capture with the estimation of opacity and transmittance, critical components of full material appearance (SVBSDF), improving the results for any material captured with a flatbed scanner, at a very high resolution and accuracy.
\end{abstract}


\end{frontmatter}

\begin{figure*}[h!]
	\centering
	\includegraphics[width=1.0\textwidth]{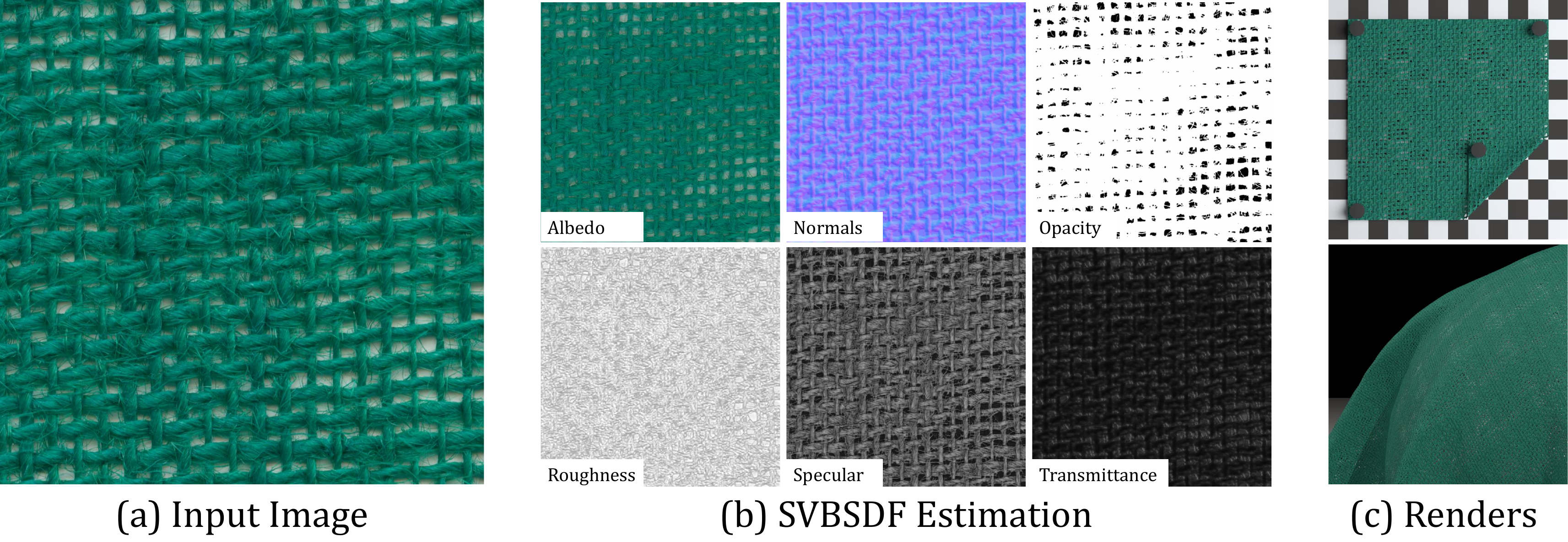}
	\vspace{0mm}
	\caption{From a single image captured with any flatbed scanner (a), our method estimates a set of high-resolution SVBSDF maps (b), which can be used in any render engine (c). }
	\label{fig:teaser}
	\vspace{-5mm}
\end{figure*}

\section{Introduction}
\label{sec:introduction}

Several industries, such as architectural and fashion design, or media and gaming, benefit from realistic digital replicas of physical materials. Yet, crafting these copies remains a laborious and slow task, demanding skilled artists, or sophisticated and expensive hardware~\cite{garces2023towards,TAC7}. Consequently, recent research has focused on devising affordable and user-friendly capture setups.

\begin{figure*}[tb!]
	\centering
	\includegraphics[width=1.0\textwidth]{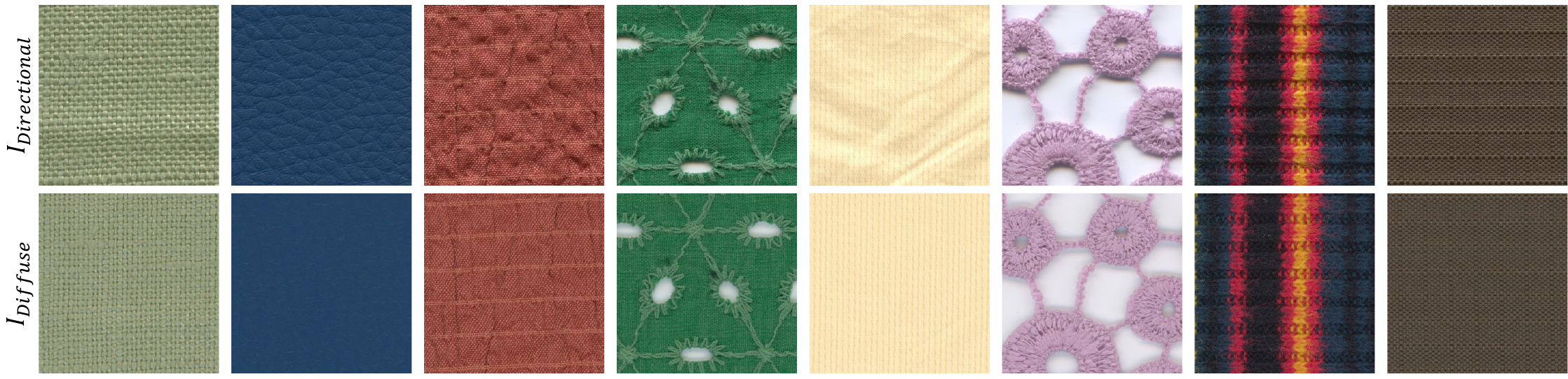}
	\caption{Some materials in our test dataset, captured on the same flatbed scanner using directional and diffuse illuminations, better suited for material capture. }
	\label{fig:dataset_delighting}
\end{figure*}

In this scenario, flatbed scanners have emerged as promising tools for high-resolution material capture~\cite{rodriguezpardo2023UMat}, owing to their user-friendly nature and provision of uniform illumination conditions. High-end scanners can even offer a lighting type closely resembling diffuse illumination, usable directly as an albedo image~\cite{rodriguezpardo2023UMat}. Nevertheless, most scanners lack this functionality, with a majority featuring a single directional light that leads to undesirable micro-specular reflections, directional shading, and cast shadows (depicted in Figure~\ref{fig:teaser}(a) and \ref{fig:dataset_delighting}).

In this work, we address the drawbacks of prior approaches and introduce a technique for digitizing materials using any scanner, removing undesirable shading and specular highlights. We show that the na\"ive solution employing an image-to-image translation network~\cite{martin2019lighting, rodriguezpardo2023UMat} falls short for this purpose. Instead, we suggest employing a cycle-consistency loss in combination with a residual formulation inspired by intrinsic image decomposition methods~\cite{garces2022survey}.

In addition, a key contribution of our method is to expand the realism of the digital replica by including opacity and transmittance in the material model. These attributes are critical for thin-layer materials, like textiles, but have been neglected in current literature. We estimate the parameters of a Spatially-Varying Bidirectional Scattering Distribution Function (SVBSDF) that can reproduce complex effects of light as it passes through the material, thereby augmenting its realism in virtual environments.

We evaluate our method using extensive and thorough experiments, leveraging image-based metrics that measure the precision of each map individually, and render-aware metrics that measure the final appearance of the material in a global context. We further demonstrate that our method works with a variety of scanning devices, producing effective results even with less controllable devices such as smartphones.

\section{Related Work}

\label{sec:related_work}

\paragraph*{\textbf{Single-Image Material Capture}} Estimating full reflectance properties of a material, using only a single image of it, is a challenging problem which has been tackled extensively in the literature in the recent years. These approaches can be categorized based on the estimation method, the imaging device employed, and the range of digitizable reflectance properties.

Neural style transfer~\cite{gatys2015neural} can be leveraged for single-image capture of stochastic materials, by matching the latent statistics of input images and renders of estimated SVBRDFs~\cite{henzler2021neuralmaterial, aittala2016reflectance}. A more common approach is to train an image-to-image translation model which takes a single image as input and estimates the set of SVBRDF maps. Originally supervised using pixel-wise or render-aware losses~\cite{deschaintre2018single,li2018materials,ye2018single,gao2019deep}, these methods have been improved by incorporating cascaded estimation~\cite{li2018learning, sang2020single}, adversarial losses~\cite{rodriguezpardo2023UMat, wen2022svbrdf, guo2021highlight,zhou2021adversarial,zhou2022tilegen,vecchio2021surfacenet},  inference-time optimization~\cite{gao2019deep}, or refinement~\cite{luo2024single}. More recently, diffusion models have emerged as powerful material estimators, showing competitive results~\cite{vecchio2023controlmat,vecchio2023matfuse,yuan2024diffmat,Sartor:2023:MFA}. A complementary line of work uses procedural graphs for material estimation~\cite{shi2020match, hu2022inverse, guo2020bayesian,jin2022woven}.

In terms of devices, the most common setup encompasses fronto-planar flash-lit images captured with a smartphone. Other setups trade this simplicity for quality, such as LCD screens~\cite{aittala2013practical,zhang2023deep,xu2023unified}, or high-end flatbed scanners~\cite{rodriguezpardo2023UMat}. Single-image material estimation methods typically estimate a reduced number of SVBRDF parameters, with the exception of~\cite{vecchio2023controlmat}, which also estimates opacity. 

Our approach differs from previous work in two ways. First, we provide a generic framework for material capture from~\emph{any} flatbed scanner, with arbitrary directional illumination, effectively removing the limitations in~\cite{rodriguezpardo2023UMat}. Furthermore, to the best of our knowledge, our method is the first single-image material capture which can estimate a full SVBSDF of a material, incorporating important effects like transmittance and opacity while preserving a high level of accuracy and resolution.

\paragraph*{\textbf{Material Delighting}} Removing shading and specular highlights from images has been explored extensively in the literature, with particular focus on removing strong shadows~\cite{qu2017deshadownet,Vasluianu_2023_CVPR,Fu_2021_CVPR,Wang_2018_CVPR} and human relighting~\cite{lagunas2021single,yeh2022learning,wimbauer2022rendering,ji2022geometry}. In the context of BRDF estimation, material delighting has been explored by combining convolutional neural networks with Poisson optimization~\cite{martin2019lighting}. Our method also leverages material delighting for albedo estimation, by incorporating ideas from intrinsic image decomposition~\cite{garces2022survey}.

\paragraph*{\textbf{Cycle-Consistent Generative Models}} Learning to map from two distinct image domains for image-to-image translation tasks can be tackled through cycle-consistent generative models~\cite{zhu2017unpaired}. By introducing the cycle-consistency loss, these models enable accurate and diverse mappings. These have shown impressive results on a wide variety of applications, including stenography~\cite{chu2017cyclegan}, voice conversion~\cite{kaneko2019cyclegan}, medical imaging~\cite{yang2020unsupervised,harms2019paired}, face generation~\cite{lu2018attribute}, or improving diffusion models~\cite{wu2022unifying,su2022dual}. Inspired by these methods, we leverage cycle-consistency to train a model capable of both material delighting and relighting, which showcases high accuracy in both tasks under several metrics.

\subsection{\textbf{Preliminaries: Material Model}}
Building upon previous work~\cite{rodriguezpardo2023UMat}, we use a physically-based material model based on microfacets reflectance~\cite{burley2012physically}, into which we incorporate additional parameters to enable transmittance effects. Our material model aggregates a diffuse component (i.e. the material albedo) $\mathbf{A} \in \mathbb{R}^{3 \times x  \times y}$, with a grayscale, isotropic specular GGX~\cite{walter2007microfacet} lobe $s_{l,v} \in \mathbb{R}^{x  \times y}$, which depends on the surface normal $\mathbf{N}$, its specularity $\mathbf{S}$ and roughness $\mathbf{R}$. The shading model $f_{l,v}^\textrm{BSDF} \in \mathbb{R}^{4 \times x  \times y}$ for a particular light $l$ and camera $v$ has an additional transparency term which depends on the material binary opacity $\mathbf{O} \in \mathbb{Z}_2^{x  \times y}$ and its transmittance  $\mathbf{T} \in \mathbb{R}^{x  \times y}$ , as follows: 

\begin{align}
	\label{eq:mat_model}
		f_{l,v}^\textrm{BSDF} (\mathbf{A},\mathbf{N},\mathbf{S},\mathbf{R},\mathbf{O},\mathbf{T})   = \mathbf{O}\cdot( \underbrace{\frac{\mathbf{A}}{\pi} + s_{l,v}(\mathbf{N},\mathbf{S},\mathbf{R})}_\text{reflectance $f_{l,v}^\textrm{BRDF}$}+ \underbrace{(\mathbf{T}\cdot\mathbf{A}))}_\text{transmittance}%
\end{align}

The transmittance is modeled as the base albedo $\mathbf{A}$ modulated by a  gray scale value $\mathbf{T}$. This assumes that the light scattered through the material is a linear attenuation of the reflectance wavelength (albedo). Finally, both reflectance and transmission are weighted by the binary operator $\mathbf{O}$, which differentiates areas with partial and total transmission.
Finally, both reflectance and transmission are weighted by the binary operator $\mathbf{O}$, which differentiates areas with partial transmission from fully transparent pixels. The distinction between both $\mathbf{O}$ and $\mathbf{T}$, being the former just a particular threshold on the continuous transmittance $\mathbf{T}$, is due to its traditional use as a binary mask in several rendering methods, to reduce shader execution time by discarding pixels.

Although there are richer and more complex models for transmittance and sub-surface scattering phenomena (E.g.:~\cite{burley2015extending}), we find that this thin-layer diffuse transmission model suffices to represent a large proportion of materials that can be captured with a scanner, while having low requirements for real time visualization and less memory consumption that a multi-channel transmission map. 

\section{Method}
\label{sec:method}

Our method takes as input a single image of the material and estimates its spatially-varying SVBSDF material parameters, including reflection and transmission per-pixel coefficients. 
The input image can be obtained with any capture device that provides mostly \textit{uniform} lighting, such as the one provided by flatbed scanners. %
Our algorithm has two steps. 
In the first step, described in Section~\ref{sec:delighting}, we use a  cycle-consistent residual generative network to \textit{delight} the material and obtain an albedo-like reflectance map. After our processing, the resulting map lacks micro-reflections and shadows that might be originally present due to directional lighting hitting the material.
In the second step, described in Section~\ref{sec:svbsdf_estimation}, we use this image as input of an attention-guided U-Net that estimate the remaining material maps, to convey reflection and transmission. 
\begin{figure*}[t!]
	\centering
	\includegraphics[width=1.0\textwidth]{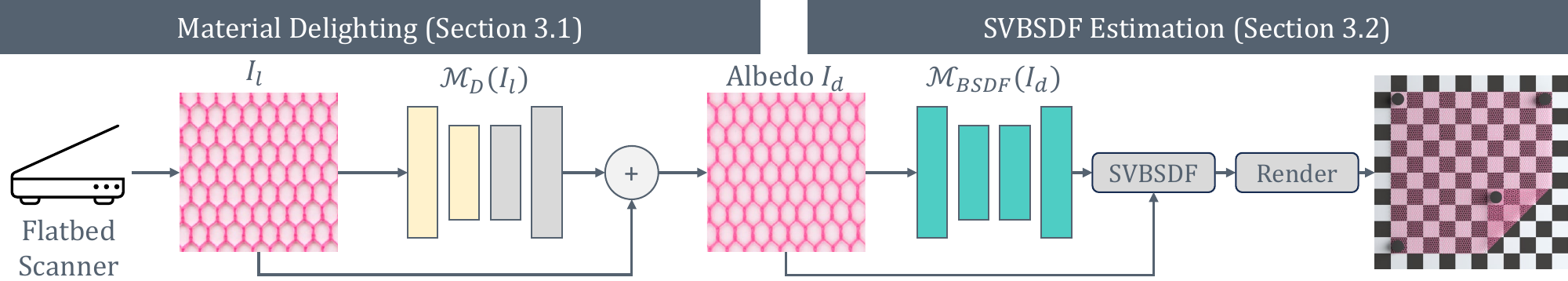}
	\caption{From an image $I_{l}$ captured with any flatbed scanner, we first estimate its albedo $I_d$ using a residual generative model $\mathcal{M}_{D}$, which removes specular highlights and shading. Taking $I_d$ as input, a second model $\mathcal{M}_{BSDF}$ estimates the rest of the SVBSDF, namely the surface normals, roughness, specular, transmittance, and opacity maps. These can be then rendered to generate photo-realistic images.    }
	\label{fig:method}
\end{figure*}

\label{sec:dataset}

\subsection{\textbf{Material Delighting}}
\label{sec:delighting}

\begin{figure*}[tb]
	\centering
	\includegraphics[width=1.0\textwidth]{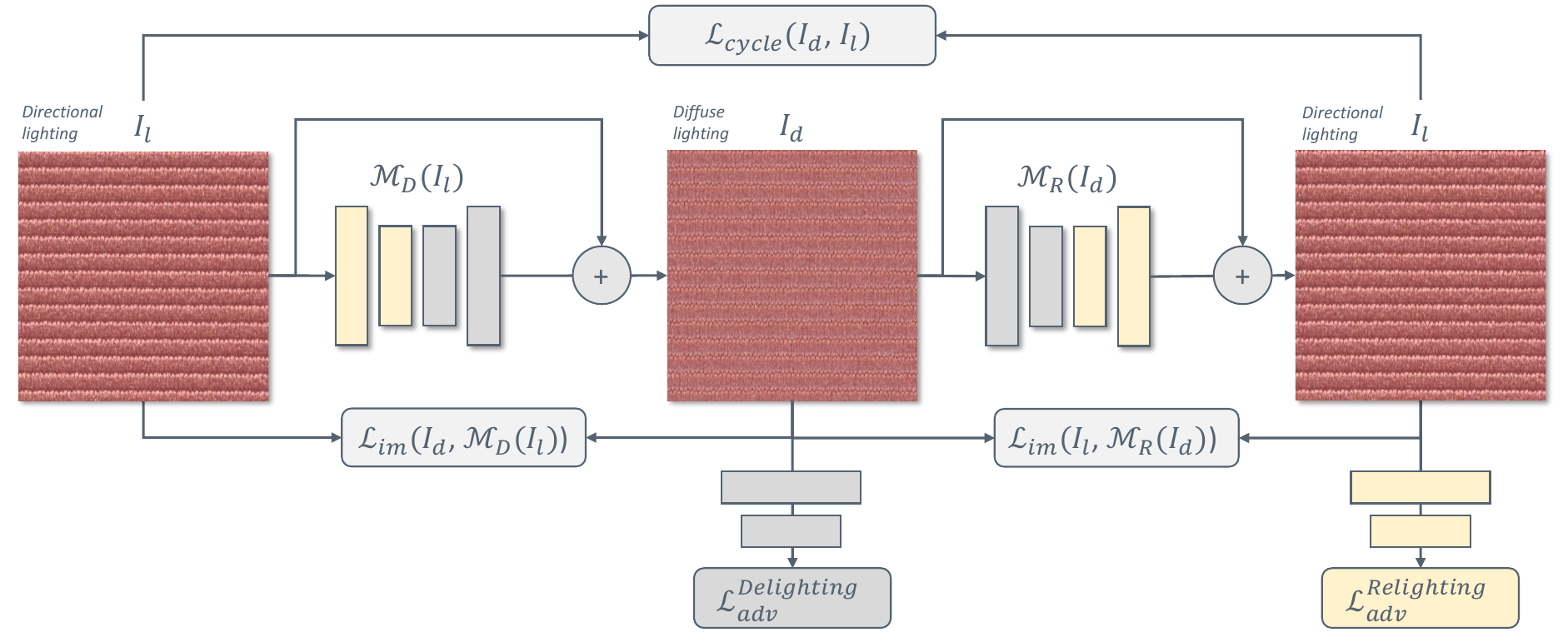}
	\caption{Diagram of our cycle-consistent generative model capable of both material delighting and relighting. }
	\label{fig:delighting_diagram}
\end{figure*}

In this step, our goal is to estimate an albedo-like reflectance map $I_{d} \approx \mathcal{A}$ from a single image $I_{l}$ of the material taken under any kind of uniform lighting. 
We term this process \textit{delighting}, as we aim to remove specular reflections, shadings, or shadows. 
A straightforward solution to this problem would be to train an image-to-image translation approach with labeled data. However, as we demonstrate, this baseline approach does not achieve the desired level of accuracy due to the under-constrained nature of the problem and our relatively reduced training dataset (see Table \ref{tab:delighting_quantitative}). Therefore, we propose a more sophisticated architecture to improve this performance, which uses residual learning and a cycle-consistency loss. 
Inspired by intrinsic image decomposition~\cite{garces2022survey}, we formulate the \textit{delighting} problem as estimating a residual layer $\mathcal{M}_R$ that adds to the albedo image to form a \textit{lighted} image, $I_{l} = I_{d} + \mathcal{M}_R (I_{d})$. Similarly, within our cycle-consistent architecture, the equivalent inverse operation also exists, and we term it \textit{relighting}, $I_{d} = I_{l} + \mathcal{M}_D (I_{l})$.
Our residuals $\mathcal{M}_R(I_{d})$ and $\mathcal{M}_D(I_{l})$ are RGB images to make the estimation more flexible, thereby removing the assumption that either the source or reflected lights are white. Figure~\ref{fig:delighting_diagram} presents an overview of the architecture.

\paragraph*{\textbf{Loss Function}}

Our loss for each branch of our cycle-consistency model is a combination of pixel-wise, perceptual, frequency, and adversarial losses,
\begin{align} 
	\Loss_\textrm{im} (\cdot, \cdot) &= \lambda_1\Loss_1 (\cdot, \cdot) +  \lambda_{\Loss_{perc}}\Loss_{perc} (\cdot, \cdot) + \lambda_{\Loss_{freq}}\Loss_{freq} (\cdot, \cdot)  + \lambda_{adv}\Loss_{adv}.
\end{align}
Following~\cite{rodriguezpardo2023UMat,rodriguez2023neubtf,garces2023towards}, for $\mathcal{L}_{perc}$ we use the AlexNet version of~\cite{zhang2018unreasonable} and for $\mathcal{L}_{freq}$ we measure the Focal Frequency Loss~\cite{jiang2021focal}. For the adversarial loss, we follow the methodology specified in~\cite{zhu2017unpaired}. Then, we build our cycle-consistency loss and full loss as,
\begin{align}
	\Loss_\textrm{cycle} (I_d, I_l) &= \underbrace{\Loss_\textrm{im}(I_d, \mathcal{M}_D(\mathcal{M}_R(I_d)))}_\text{delighting} + \underbrace{\Loss_\textrm{im}(I_l, \mathcal{M}_R(\mathcal{M}_D(I_l))}_\text{relighting}	\\
	\Loss_\textrm{full}  (I_d, I_l) &= \Loss_\textrm{im} (I_d, \mathcal{M}_D(I_{l}) ) +  \Loss_\textrm{im} (I_l, \mathcal{M}_R(I_{d}) )  + \lambda_{cycle}\Loss_{\textrm{cycle}} (I_d, I_l).
\end{align}

\REMOVE{
an image captured with any flatbed scanner $I_{l}$, which may contain specular reflections, shadings or shadows. We formulate this problem with an image-to-image translation approach, leveraging an attention-guided convolutional U-Net~\cite{ronneberger2015u} $\hat{A} = \mathcal{M}_D (I_{l})$. 

To train this model, we leverage our dataset described in Section~\ref{sec:dataset}, which contains pairs of images $I_{d}$ and $I_{l}$, which are pixel-wise coherent. While directly training a model which maps from $I_{l}$ to $I_{d}$ using a standard pixel-wise norm could yield acceptable results, we empirically observe that this baseline approach does not achieve the desired level of accuracy. Therefore, we introduce a series of modifications aimed at maximizing generalization. 

First, inspired by intrinsic image decomposition~\cite{garces2022survey}, we formulate this mapping as learning an residual shading component, effectively learning a relationship $I_{d} = I_{l} + \mathcal{M}_D (I_{l})$. This simple modification allows for better training dynamics and generalization capabilities. Note that we use an RGB residual to make the estimation more flexible, thereby removing the assumption that either the source or reflected lights are white.

\paragraph*{Cycle-Consistency}  
Importantly, in parallel to $M_D$, we train a \emph{relighting} neural network which learns to estimate $I_{l} = I_{d} + \mathcal{M}_R (I_{d})$, and use its estimations to further guide the training of $\mathcal{M}_D$, which is our main objective. We therefore formulate this learning problem as a cycle-consistent generative adversarial network, in which both delighting and relighting models are supervised with ground truth data, and with each other's estimations. This increases model accuracy and robustness, effectively working as a data-augmentation policy.

\paragraph*{Loss Function}

Our loss for the task of model delighting is a combination of pixel-wise, perceptual, frequency and adversarial losses, as follows:

\begin{align} \label{eq:delighting}
	\resizebox{\hsize}{!}{$\mathcal{L}_{\text{\tiny{Delighting}}} (I_d, I_l) = \lambda_{\mathcal{L}_1}\mathcal{L}_1^D (\mathcal{M}_D(I_{l}), I_{d}) +  \lambda_{\mathcal{L}_{perc}}\mathcal{L}_{perc}^D (\mathcal{M}_D(I_{l}), I_{d}) + \lambda_{\mathcal{L}_{freq}}\mathcal{L}_{freq}^D (\mathcal{M}_D(I_{d}), I_{l})  + \lambda_{adv}\mathcal{L}_{adv}^D $}
\end{align}

Following~\cite{rodriguezpardo2023UMat,rodriguez2023neubtf,garces2023towards}, for $\mathcal{L}_{perc}$ we use the AlexNet version of~\cite{zhang2018unreasonable} and for $\mathcal{L}_{freq}$ we measure the Focal Frequency Loss~\cite{jiang2021focal}. For the adversarial loss, we follow the methodology specified in~\cite{zhu2017unpaired}. Similarly, our loss for the relighting model is:

\begin{align} \label{eq:relighting}
	\resizebox{\hsize}{!}{$\mathcal{L}_{\text{\tiny{Relighting}}} (I_d, I_l)= \lambda_{\mathcal{L}_1}\mathcal{L}_1^R (\mathcal{M}_R(I_{d}), I_{l}) +  \lambda_{\mathcal{L}_{perc}}\mathcal{L}_{perc}^R (\mathcal{M}_R(I_{d}), I_{l}) + \lambda_{\mathcal{L}_{freq}}\mathcal{L}_{freq}^R (\mathcal{M}_R(I_{l}), I_{d})  + \lambda_{adv}\mathcal{L}_{adv}^R $}
\end{align}

Our cycle consistency loss is defined as:
\begin{align} \label{eq:cycle_loss}
	\resizebox{\hsize}{!}{$\mathcal{L}_{\text{\tiny{cycle}}} (I_d, I_l) = \mathcal{L}_{\text{\tiny{Relighting}}}(I_d, \mathcal{M}_R(\mathcal{M}_D(I_l))) + \mathcal{L}_{\text{\tiny{Delighting}}}(I_l, \mathcal{M}_D(\mathcal{M}_R(I_d)))           $} 
\end{align}

The final loss is defined as follows:

\begin{align} \label{eq:cycle_loss}
	\resizebox{.7\hsize}{!}{$\mathcal{L} = \mathcal{L}_{\text{\tiny{Delighting}}} +  \mathcal{L}_{\text{\tiny{Relighting}}}  + \lambda_{cycle}\mathcal{L}_{\text{\tiny{cycle}}}                    $} 
\end{align}

}

\paragraph*{\textbf{Architecture Design}} For the generator architectures, we follow the attention-guided U-Net design in~\cite{rodriguezpardo2023UMat}, using a single decoder for each model, and removing the MLPs appended to the end of the architecture. For the discriminators, we follow previous work on texture synthesis~\cite{rodriguezpardo2022seamlessgan,zhou2018non} and use a 4-layer PatchGAN~\cite{isola2017image}. 

\paragraph*{\textbf{Data Augmentation}} We train the models using random cropping, with $128\times128$ resolution patches. Besides, we use random rescaling, enabling the model to generalize on the $(300, 1200)$ PPI range, covering most flatbed scanners. Finally, we use random horizontal and vertical flips to further enhance generalization. 

\paragraph*{\textbf{Implementation Details}} We standardize each dataset of directional and diffuse images using their respective means and standard deviations, enabling the model to focus on relative differences and not on global average values. We use PyTorch~\cite{paszke2017automatic} and Torchvision~\cite{marcel2010torchvision} for training, and Kornia~\cite{riba2020kornia} for data augmentation. We train the models using Adam~\cite{kingma2014adam} for 100 iterations, with an initial learning rate of $0.002$, halved every 30 iterations. We leverage automatic gradient scaling and mixed precision training~\cite{micikevicius2017mixed}. Both generators and discriminators are initialized using \emph{orthogonal initialization}~\cite{hu2020provable}. Training these models takes approximately 12 hours on a NVidia 3060 GPU. After a Bayesian hyperparameter~\cite{wandb} optimization performed on a separate validation dataset, we set the loss weighting as $\lambda_{cycle} = 0.25, \lambda_{adv} = 0.15, \lambda_{perc} = 0.3, \lambda_{freq} = 0.2, \lambda_{\mathcal{L}_1} = 1$. Further details are included in the supplementary material. 

\subsection{\textbf{SVBSDF Estimation}}
\label{sec:svbsdf_estimation}

To estimate the rest of the SVBSDF, we build upon the training methodology described in~\cite{rodriguezpardo2023UMat}. First, we expand their model to enable the estimation of opacity and transmittance maps, thus introducing two additional decoders to their attention-guided U-Net network, and expanding the loss function and discriminator architecture accordingly. 

We further introduce additional minor changes to improve the estimation. Most notably, we parameterize the normal map so as to estimate $\theta, \phi$ angles instead of full cartesian coordinates $xyz$, and following~\cite{garces2023towards} adopt~\emph{elliptical grid mapping}~\cite{fong2015analytical} for additional performance gains. %
Finally, we use AdamW~\cite{loshchilov2017decoupled} and $256\times256$ crops for training, and perform minor hyperparameter changes, which are fully described in the supplementary material.

\section{Evaluation}

\subsection{\textbf{Dataset}}
Using a high-end \emph{EPSON V850 Pro} flatbed scanner, we capture 3830 10x10 cm material samples at 1200 PPI resolution. %
Note that this scanner can capture images using a standard, single LED strip illumination, like lower-end scanners, but also enables a higher-quality setup using a dual-light which provides diffuse-like illumination. A detailed description of the dataset is provided in the supplementary material.
The later setup closely resembles fitted albedos~\cite{rodriguezpardo2023UMat}, removing strong shades caused by wrinkles or mesostructure, hiding shadows casted to the scanner lid and eliminating specular highlights (as shown in Figure~\ref{fig:dataset_delighting}).
We thus capture two images for each material: $I_{l}$ and $I_{d}$, preserving pixel-wise correspondence for every material under both illuminations. To augment this dataset, we further capture every material on their front and back sides, and rotate them by 90º to allow for generalization to multiple orientations. We also capture these materials on a custom gonioreflectometer, and leverage the methodology described in~\cite{rodriguezpardo2021transfer, garces2023towards} to propagate the ground truth material parameters described in Equation~\ref{eq:mat_model}. 
We use 10\% of this dataset for testing.

\subsection{\textbf{Metrics}}
To measure the performance of our models, we use a variety of metrics aimed at understanding the perceptual, pixel-wise, and render-aware accuracy of our estimations.
First, to measure the errors of our generators, $\mathcal{M}_D$ and  $\mathcal{M}_R$, we leverage traditional image quality metrics, as well as $\Delta E$~\cite{mokrzycki2011colour}, which accurately measures color differences, and perceptually-motivated alternatives like FLIP~\cite{Andersson2020} and LPIPS~\cite{zhang2018unreasonable}. We also quantify per-map accuracy, leveraging pixel-wise $\mathcal{L}_1$ norms for the Albedo, Roughness, Specular, and Transmittance maps, angular distances $\mathcal{L}_\measuredangle$ for the surface normals, and the Jaccard index $\mathcal{L}_{Jacc}$ for the opacity maps. Following~\cite{rodriguezpardo2023UMat}, we also report Pearson correlations $\rho$.

The previous metrics are useful to assess individual precision of the estimations. However, when reproducing a real material, it is of critical  importance to understand how these parameters interact with each other in the integrated physically-based rendering space. Thus, we propose a set of metrics aimed to evaluate the accuracy of the full material model in terms of both reflectance and transmittance. 
For reflectance, we expand the $\mathcal{L}_{\text{\tiny{BRDF}}}$ metric in~\cite{rodriguezpardo2023UMat}, which measures the perceptual error, with extra terms that account for material opacity, and cosine weighting and peak reflectance attenuation to account for human visual perception~\cite{lavoue2021perceptual}. 
We measure the render-space reflectance estimation difference between the ground truth $\mathbf{M}_{GT}$ and predicted $\hat{\mathbf{M}}$ material as follows:
\begin{align} \label{eq:brdf}
	\resizebox{\hsize}{!}{$\mathcal{L}_{\text{\tiny{BRDF}}} (\mathbf{M}_{GT},\hat{\mathbf{M}}) = \frac{1}{xy} \sum\limits_{xy} \sqrt{ \frac{1}{|S|}\sum\limits_{(l, v) \in S } \sqrt[3]{\cos^2(\theta_l) \left( f_{l,v}^\textrm{BRDF}(\mathbf{A}_{GT},\mathbf{N}_{GT},\mathbf{S}_{GT},\mathbf{R}_{GT})\cdot O_{GT} - f_{l,v}^\textrm{BRDF}(\hat{\mathbf{A}},\hat{\mathbf{N}},\hat{\mathbf{S}}, \hat{\mathbf{R}}) \cdot \hat{O}\right)^2}}$}
\end{align}
where $l,v$ are a set of 50 lights and viewing angles optimized for BRDF acquisition, gathered from~\cite{nielsen2015optimal}. 

For transmittance, we introduce a novel metric, $\mathcal{L}_{\text{\tiny{BTDF}}}$, which explicitly measures the error in the estimation of transmissive effects as follows: 
\begin{align}
\label{ec:ell_svbrdf}
       \mathcal{L}_{\text{\tiny{BTDF}}} (\mathbf{M}_{GT},\hat{\mathbf{M}})= \frac{1}{xy} \sum\limits_{xy} |  \mathbf{T}_{GT} \cdot \mathbf{A}_{GT} \cdot \mathbf{O}_{GT}  - \hat{\mathbf{T}} \cdot \hat{\mathbf{A}} \cdot \hat{\mathbf{O}} | 
\end{align}
Finally, we define our final metric $\mathcal{L}_{\text{\tiny{BSDF}}}$ as a weighted combination of $\mathcal{L}_{\text{\tiny{BRDF}}}$ and $\mathcal{L}_{\text{\tiny{BTDF}}}$, setting $w_{\text{\tiny{BRDF}}}  = \frac{1}{2}$ for simplicity: 
\begin{align}
\label{ec:ell_svbsdf}
       \mathcal{L}_{\text{\tiny{BSDF}}} = w_{\text{\tiny{BRDF}}} \mathcal{L}_{\text{\tiny{BRDF}}} +  (1-w_{\text{\tiny{BRDF}}}) \mathcal{L}_{\text{\tiny{BTDF}}}  
\end{align}

This integrated metric is render-aware and perceptually validated and enables the comparison of different configurations of our models. 

\subsection{\textbf{Ablation Study}}

In this section, we present an ablation study to validate each of our components.

\subsubsection*{\textbf{Delighting Model}} %
Table~\ref{tab:delighting_quantitative} presents the results of the study for our \textit{Relighting} and \textit{Delighting} flows. 
Our baseline is a pure regression-based model which uses only pixel-wise $\Loss_1$ losses. We progressively add components to this baseline, to study their impact.
First, making the models generative by introducing $\Loss_{adv}$ to their training losses enables higher accuracy.
Training $\mathcal{M}_D$ and $\mathcal{M}_R$ together, using our cycle-consistency loss $\Loss_{cycle}$, strongly improves accuracy across every metric, providing evidence that this is a key component for achieving high-quality albedo estimations. 
Further, using our residual learning approach, inspired by intrinsic decomposition, provides significant gains across every metric.
Finally, incremental improvements are achieved by introducing $\Loss_{perc}$ and $\Loss_{freq}$, and our full data augmentation policy. Interestingly, we observe that relighting is typically a harder task.
We believe that our cycle-consistent approach allows our model to generalize better because it works as a form of data augmentation, while residual learning improves training dynamics and makes the task easier to learn.

\begin{table*}[t]
\centering
\resizebox{\textwidth}{!}{%
\begin{tabular}{@{}lccccc|ccccc@{}}
\cmidrule(l){2-11}
\multicolumn{1}{c}{\multirow{2}{*}{}}      & \multicolumn{5}{c}{\textbf{Relighting}} & \multicolumn{5}{c}{\textbf{Delighting}} \\ \cmidrule(l){2-11} 
 &
 $        $ PSNR $\uparrow$ &
  SSIM~\cite{wang2004image} $\uparrow$ &
  LPIPS~\cite{zhang2018unreasonable} $\downarrow$   &
  $\Delta E $ $\downarrow$ &
  \FLIP~\cite{Andersson2020} $\downarrow$ &
  $        $ PSNR $\uparrow$&
  SSIM~\cite{wang2004image} $\uparrow$ &
  LPIPS~\cite{zhang2018unreasonable} $\downarrow$   &
  $\Delta E $ $\downarrow$ &
  \FLIP~\cite{Andersson2020} $\downarrow$ \\ \cmidrule(l){2-11} 
  \multicolumn{1}{l}{No Delighting}    &  24.01    & 0.686    &  0.262   &  4.732    &  0.264   &   24.01    & 0.686    &  0.262   &  4.732    &  0.264    \\ \cmidrule(l){1-11} 
\multicolumn{1}{l}{Baseline Delighting}              & \RedColor{24.47}     &  \RedColor{0.784}    & \RedColor{0.289}     &  \RedColor{6.102}    &  \RedColor{0.259}   &  \RedColor{25.71}    &  \RedColor{0.798}    & \RedColor{0.267}    &  \RedColor{4.922}    &  \RedColor{0.219}    \\
\multicolumn{1}{l}{+ $\mathcal{L}_{adv}$}    & 24.69     &  0.809    &  0.257    & 5.587     &  0.244   &   26.72   &  0.810    &  0.242     &   4.359  &  0.202   \\
\multicolumn{1}{l}{+ Cycle-Consistency}   &  26.54    &  0.856    & 0.218     &  4.213    & 0.194    &  27.82    &  0.851    &  0.202    &  3.604   &  0.169   \\
\multicolumn{1}{l}{+ Residual}            & 28.42     & 0.903     &  0.165    & 3.115    & 0.147    &  29.79    &  0.897    &  0.171    &  2.754   &   0.132  \\
\multicolumn{1}{l}{+ Full Loss}           &   \GreenColor{28.67}   &  \GreenColor{0.912}   &  0.164    &   3.071   &  0.144   &   30.19   &   0.906   &   0.151   & 2.630    &  0.126   \\
\multicolumn{1}{l}{+ Aug. (Final Model)}     &  28.48    & 0.907     &  \GreenColor{0.161}    &  \GreenColor{3.012}    & \GreenColor{0.138}    & \GreenColor{31.41}    &  \GreenColor{0.933}    &  \GreenColor{0.136}    &  \GreenColor{2.261}   &   \GreenColor{0.111}  \\ \bottomrule
\end{tabular}%
}
\caption{Results of our ablation study of our material delighting algorithm, across a variety of metrics. On the top row, we show the results when no delighting is applied. We use a color code to highlight \GreenColor{best} and \RedColor{worst} cases.}
\label{tab:delighting_quantitative}
\end{table*}

\begin{table*}[h!]
	\centering
	\resizebox{\textwidth}{!}{%
		\begin{tabular}{@{}lcccccccccccc@{}}
			\cmidrule(l){2-13}
			& \multicolumn{6}{c}{\textbf{Pixel-Wise Errors}} & \multicolumn{3}{c}{\textbf{Correlations}} & \multicolumn{3}{c}{\textbf{Render-Aware}} \\ \cmidrule(l){2-13} 
			\multicolumn{1}{c}{} &
			\multicolumn{1}{c}{$\mathcal{L}_{1}^{A}\downarrow$} &
			\multicolumn{1}{c}{$\mathcal{L}_{\measuredangle} \downarrow$} &
			\multicolumn{1}{c}{$\mathcal{L}_{1}^{R}\downarrow$} &
			\multicolumn{1}{c}{$\mathcal{L}_{1}^{S}\downarrow$} &
			\multicolumn{1}{c}{$\mathcal{L}_{1}^{T}\downarrow$} &
			\multicolumn{1}{c}{$\mathcal{L}_{Jac}^{O}\uparrow$} &
			$\rho^{R}\uparrow$ &
			$\rho^{S}\uparrow$ &
			$\rho^{T}\uparrow$ &
			\multicolumn{1}{c}{$\mathcal{L}_{BRDF} \downarrow$} &
			\multicolumn{1}{c}{$\mathcal{L}_{BTDF} \downarrow$} &
			\multicolumn{1}{c}{$\mathcal{L}_{BSDF} \downarrow$} \\ \cmidrule(l){2-13} 
			\multicolumn{1}{l}{\textbf{UMat}~\cite{rodriguezpardo2023UMat}, Diffuse Illumination \dag}              & 0.000*    & 2.666 & 0.060 & 0.086 &  0.073 & \multicolumn{1}{c|}{0.931} & 0.714   &   0.852  & \multicolumn{1}{c|}{0.000}   &   0.324        &  0.058         &    0.191     \\ 
			\multicolumn{1}{l}{\textbf{UMat}~\cite{rodriguezpardo2023UMat}, Directional Illumination \dag}              & 0.051    & 2.813 &  0.062  & 0.089 & 0.073 & \multicolumn{1}{c|}{0.931} & 0.701   &   0.841  & \multicolumn{1}{c|}{0.000}  &  0.384       &  0.089          &  0.237     \\ \cmidrule(l){1-13} 
			\multicolumn{1}{l}{\textbf{Ours}, w/o Delighting} \\
			\multicolumn{1}{l}{\textbf{} Diffuse Illumination}              & 0.000*  & 2.221   & 0.055  & 0.081 & 0.017 & \multicolumn{1}{c|}{0.998} &  0.731   &   0.899   & \multicolumn{1}{c|}{0.937}  &   0.223         &    0.026       &   0.125 \\ 
			\multicolumn{1}{l}{\textbf{} Directional Illumination}              & 0.051  & 2.771  &  0.061   & 0.086  & 0.025 & \multicolumn{1}{c|}{0.958} & 0.711  & 0.877   & \multicolumn{1}{c|}{0.897}  &  0.344         &   0.059        &   0.202      \\ \cmidrule(l){1-13} 
			\multicolumn{1}{l}{\textbf{Ours} w/ Baseline Delighting}              & \RedColor{0.042} & \RedColor{3.241} & \RedColor{0.063}  & \RedColor{0.088} & \RedColor{0.035} & \multicolumn{1}{l|}{\RedColor{0.945}} & \RedColor{0.675}   & \RedColor{0.852}    & \multicolumn{1}{c|}{0.873}  &       \RedColor{0.362}     &     \RedColor{0.071}      & \RedColor{0.217}        \\
			\multicolumn{1}{l}{+ $\mathcal{L}_{adv}$} & 0.037 & 2.981 & 0.062 & 0.086  & 0.034 & \multicolumn{1}{c|}{0.952} &  0.701   &  0.877  & \multicolumn{1}{c|}{\RedColor{0.872}}  &  0.345         & 0.067          &   0.206       \\
			\multicolumn{1}{l}{+ Cycle-Consistency}   & 0.032 & 2.692 &  \GreenColor{0.058} & 0.084 & 0.028 & \multicolumn{1}{c|}{0.981} & 0.713  & 0.889     & \multicolumn{1}{c|}{0.895}  &  0.303         &  0.051         &   0.177       \\
			\multicolumn{1}{l}{+ Residual}            & 0.026  & 2.474  &  \GreenColor{0.057} & 0.086 & 0.021  & \multicolumn{1}{c|}{0.992} &  \GreenColor{0.722}   & 0.881   & \multicolumn{1}{c|}{0.921}  &    0.276       &    0.038       &    0.157      \\
			\multicolumn{1}{l}{+ Full Loss}           &  0.024 & 2.441 &  \GreenColor{0.057} &  \GreenColor{0.081} & 0.023  & \multicolumn{1}{c|}{0.991} &  \GreenColor{0.721}   & 0.898   & \multicolumn{1}{c|}{0.919}  &   0.261        &  0.035         &    0.148      \\
			\multicolumn{1}{l}{+ Aug. (Final Model)}  & \GreenColor{0.021}  &  \GreenColor{2.333} &  \GreenColor{0.057} &  \GreenColor{0.080} & \GreenColor{0.019} & \multicolumn{1}{c|}{\GreenColor{0.994}} &  \GreenColor{0.722}   &   \GreenColor{0.903}   &  \multicolumn{1}{c|}{\GreenColor{0.930}}  &  \GreenColor{0.253}          &  \GreenColor{0.030}        &   \GreenColor{0.142}       \\ \bottomrule
		\end{tabular}%
	}
	\caption{Results of previous work, and of our ablation study, on final digitization accuracy, on per-map and integrated metrics. We use a color code to highlight \GreenColor{best} and \RedColor{worst} cases. Errors marked with * correspond to input images which we assume to be the ground truth albedos, hence $\mathcal{L}_{1}^{A} = 0$. \dag Note that~\cite{rodriguezpardo2023UMat} does not estimate transmittance nor opacity, instead we assume the materials are fully opaque.  }
	\label{tab:brdf_quantitative}
\end{table*}

\subsubsection*{\textbf{SVBSDF Estimation Accuracy}}
Table~\ref{tab:brdf_quantitative} presents the errors of our end-to-end digitization pipeline, including the error of not using the delighting model (\textit{Ours w/o Delighting}) and comparison with related work UMat~\cite{rodriguezpardo2023UMat}, the main available method in previous work that uses scanners as a capture device. Note that, to date, there is no previous work that estimates transmittance images. 
We also test different inputs using images captured under Diffuse \textit{albedo-like} Illumination (therefore delighting operation would not be necessary), and images with Directional Illumination. 

As shown, our model behaves consistently better across every metric than UMat. Note that UMat does not estimate opacity nor transmittance (we set all their materials to be full opaque), which is heavily penalized by the integrated metric $\Loss_{\bsdf}$. Interestingly, the relative improvements of our model are more visible on the normal map than on the roughness or specular maps, likely due to our proposed normal reparameterization. Minor training improvements like the surface normal reparameterization, some hyperparameter changes, and a larger dataset helped push accuracy further. 

We also expand our previous ablation using render metrics. Notably, we observe that more accurate delighting does not only result in better albedo estimation, but the estimation of the remainder of the SVBSDF also becomes more precise. This is particularly visible on the surface normals, transmittance and opacity maps, while roughness or specularity estimations are generally less dependent on the delighting quality. Overall, our final model achieves the best results on every metric, with our cycle-consistency and residual approaches proving to be the most impactful components of the method.

\REMOVE{In Table~\ref{tab:brdf_quantitative}, we show a quantitative comparison between different model and input configurations, on the digitization of each texture map and with our metrics, measured across our whole test set. In the first two rows, we show the quantitative results of~\cite{rodriguezpardo2023UMat}, on directional and diffuse illumination. On the second two rows, we show the results of our model for both illumination types, with no delighting applied to the input images. As shown, our model behaves consistently better across every metric than UMat~\cite{rodriguezpardo2023UMat}. Note that UMat does not estimate opacity nor transmittance, which is heavily penalized by the integrated metrics. Interestingly, the relative improvements of our model are more visible on the normal map than on the roughness or specular maps, likely due to our proposed normal reparameterization. Minor training improvements and a larger dataset helped push accuracy further. }

\REMOVE{Besides, we also show the results of SVBSDF estimation of the material delighting models we tested in our ablation study. Notably, we observe that more accurate delighting does not only result in better albedo estimation, but the estimation of the remainder of the SVBSDF also becomes more accurate. This is particularly visible on the surface normals, transmittance and opacity maps, while roughness or specularity estimations are generally less dependent on the delighting quality. Overall, our final model achieves the best results on every metric, with our cycle-consistency and residual approaches proving to be the most impactful components of the method. }

\label{sec:evaluation}

\subsection{\textbf{Qualitative Results}}
\label{sec:results}

\begin{figure*}[tb!]
	\centering
	\includegraphics[width=1.0\textwidth]{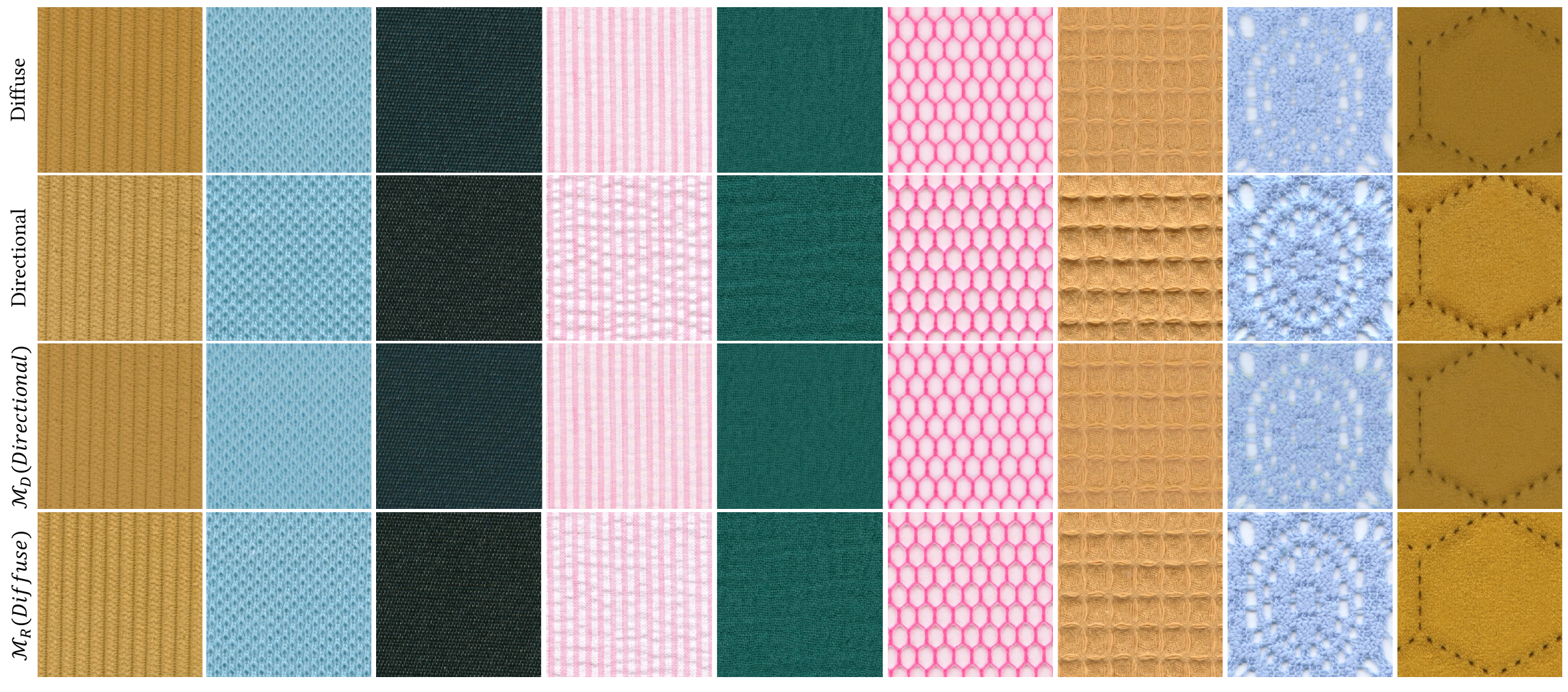}
	\caption{Qualitative results of our material delighting framework. On the first two rows, we show images captured with flatbed scanners under diffuse (top) and directional (bottom) illumination. We use those as input to our delighting $\mathcal{M}_D$ and relighting $\mathcal{M}_R$ models, respectively, for which we show the results on the bottom rows. }
	\label{fig:qualitative_delighting}
\end{figure*}

Figure~\ref{fig:qualitative_delighting} shows qualitative results of our material delighting and relighting models, along with ground truth data. Our delighting model behaves accurately even in challenging cases, like the corduroy on the first column, the satin on the third or the suede leather on the last one. The predicted images contain no shading, wrinkles are hidden and shadows casted on the scanner lid are eliminated. It can be seen why the material relighting model is less accurate according to our metrics. Precisely introducing shadows, specular highlights or shading proves to be a more challenging task than removing them, and our relighting model sometimes misplaces or inaccurately estimates the intensity of these reflections. We believe that, during training using cycle-consistency, this helps the delighting model as this works as a short of data augmentation, as the delighting model is shown variations of the same material with different variations on shading and specularity.

\begin{figure*}[tb!]
	\centering
	\includegraphics[width=1.0\textwidth]{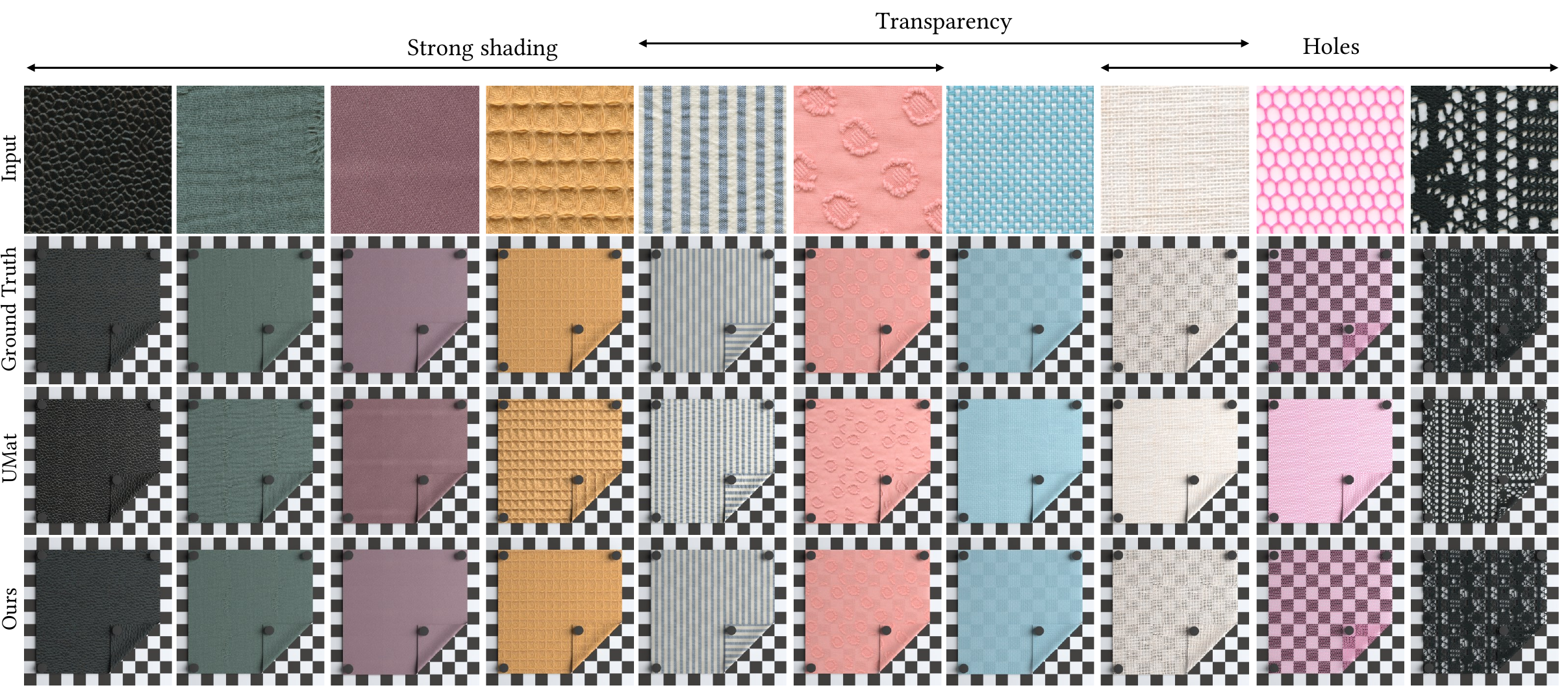}
	\caption{Qualitative comparisons of our method with UMat~\cite{rodriguezpardo2023UMat} for a few representative materials in our test set, with strong shading (leftmost columns), transparency (middle) or holes (rightmost). We show the input image (top row), and renders using the ground truth materials (captured with a gonioreflectometer), the estimation of~\cite{rodriguezpardo2023UMat} and ours, on the second, third and fourth rows, respectively. Best viewed in color on a screen.}
	\label{fig:comp_umat}
\end{figure*}

Figure~\ref{fig:comp_umat} compares our outputs with UMat~\cite{rodriguezpardo2023UMat} for a variety of material types. 
We show the results on a render scene designed to better show the impact of material transparency. As shown, our model behaves accurately across many different materials, with precise and sharp albedo estimations, and realistic transparency and opacity predictions. These results highlight the importance of estimating these maps, as the results of UMat are less appealing and realistic in comparison.

\begin{figure*}[tb!]
	\centering
	\includegraphics[width=1.0\textwidth]{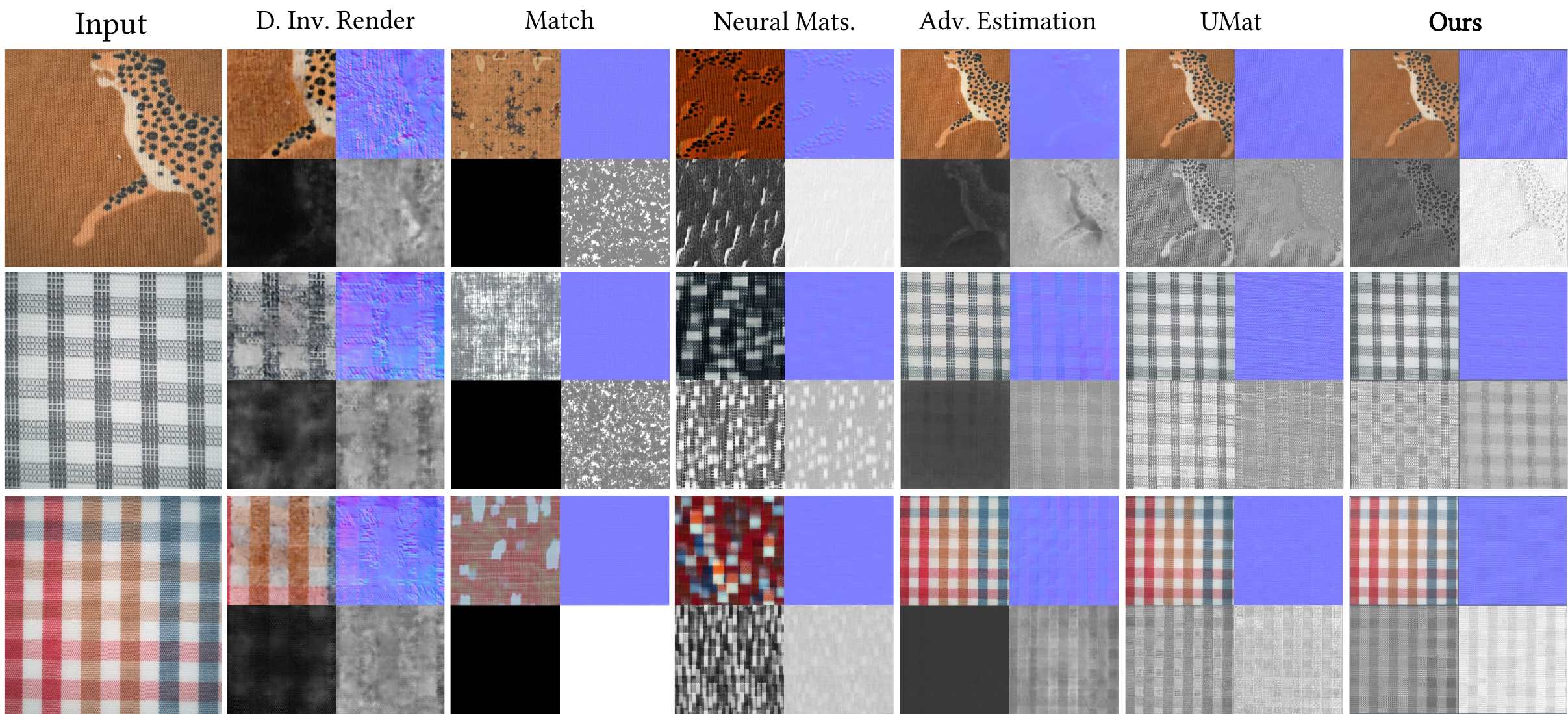}
	\caption{Comparisons of our method with previous work on images captured with a smartphone, using ambient lighting (top row) and flash illumination (bottom two rows), at different levels of resolution. From left to right, we show input images, and the results of Deep Inverse Rendering~\cite{gao2019deep}, Match~\cite{shi2020match}, Neural Materials~\cite{henzler2021neuralmaterial}, Adversarial SVBRDF Estimation~\cite{zhou2021adversarial}, UMat~\cite{rodriguezpardo2023UMat}, and ours. Note that we only show the four reflectance maps used by every method: albedo, normals, specular and roughness.}
	\label{fig:comparisons}
\end{figure*}

In Figure~\ref{fig:comparisons}, we show results of different methods of material capture, for which we capture the input images using a smartphone. We include results on ambient lighting (top row) and flash-lit images (last two rows). The methods of Deep Inverse Rendering~\cite{gao2019deep}, Match~\cite{shi2020match}, Neural Materials~\cite{henzler2021neuralmaterial}, Adversarial SVBRDF Estimation~\cite{zhou2021adversarial} all assume a smartphone capture, while UMat~\cite{rodriguezpardo2023UMat} and our method assume a flatbed scanner capture. Regardless on the illumination conditions, our model provides sharp and accurate estimations which better preserve the structure and color of the inputs compared to generative or optimization-based models~\cite{gao2019deep, shi2020match, henzler2021neuralmaterial}. Compared to UMat~\cite{rodriguezpardo2023UMat}, our delighting framework enables more uniform and higher-quality albedos, and we achieve more globally coherent maps than~\cite{zhou2021adversarial}. Overall, our method proves robust to images captured with smartphones across a variety of illumination conditions, even if we never train our models with this type of data. Note that this comparisons are done in a qualitative fashion as there is no accessible ground truth for their SVBRDF maps.

\begin{figure*}[tb!]
	\centering
	\includegraphics[width=1.0\textwidth]{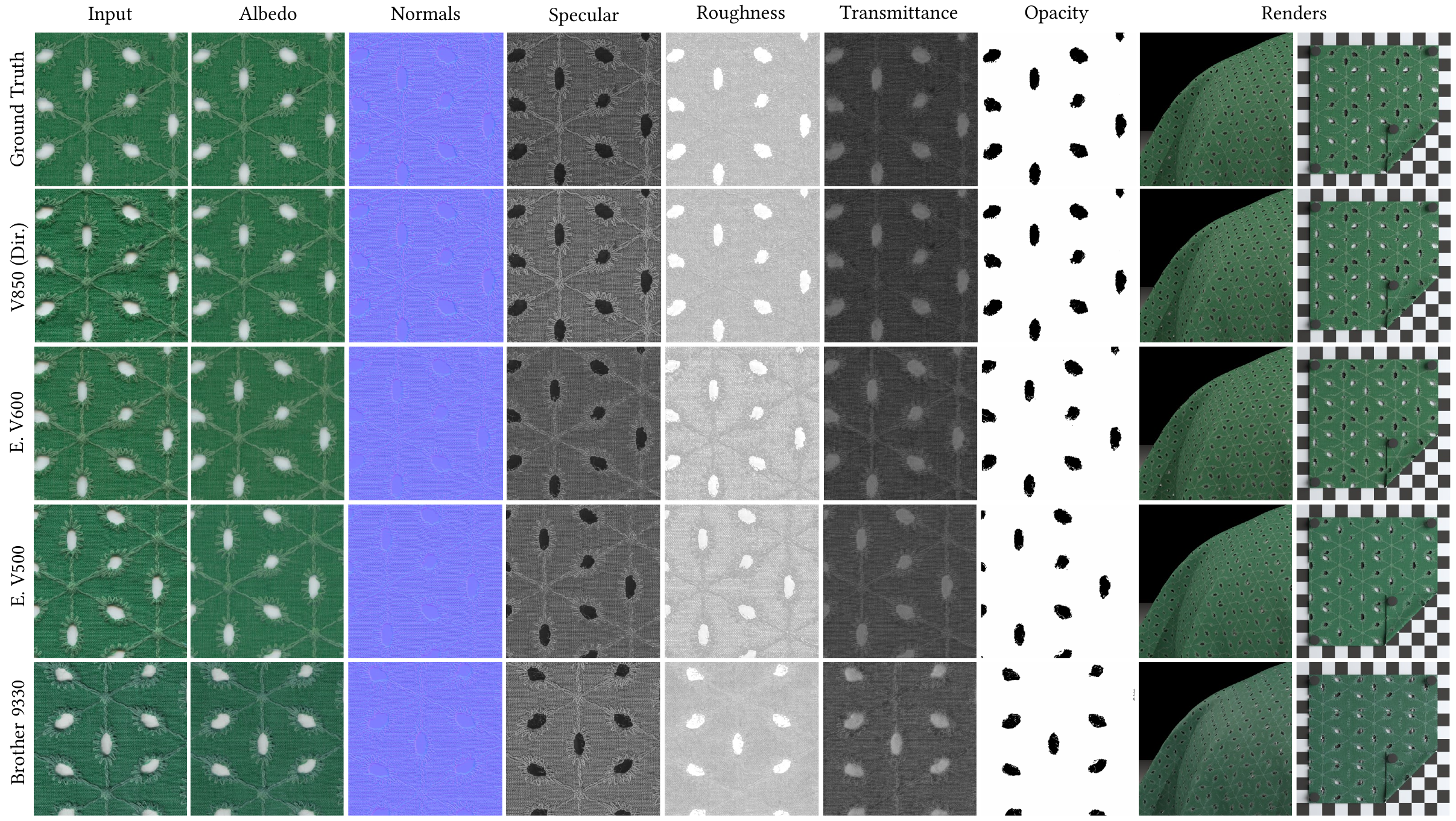}
	\caption{Qualitative comparison between several flatbed scanners for the same material. Note that the images are not exactly pixel-wise coherent across scanners. }
	\label{fig:qualitative_scanners}
\end{figure*}

In Figure~\ref{fig:qualitative_scanners}, we show the results achieved by our model on the same material, across a variety of flatbed scanners. This material is challenging, containing holes, wrinkles and fly-away fibers, all of which pose problems for digitization. The inputs on the first two rows were captured with a high-end EPSON V850 Pro scanner, for which we show the ground truth materials and renders (first), and the estimation on the directional light configuration on said scanner. On the third and fourth rows, we show the results on lower-end Epson V600 and V500 flatbed scanners, and the final row was captured with a Brother 9930 multi-functional fax machine, which also contains a budget scanner. As shown, our model estimations remain consistent across devices, regardless on the quality of the input scanner. The results for Brother 9930 struggle in terms of color due to calibration issues, but our SVBSDF estimation contains sharp, accurate reflectance maps. More results are shown in the supplementary material.

\subsection{\textbf{Failure Cases and Limitations}} 

\begin{figure*}[tb!]
	\centering
	\includegraphics[width=1.0\textwidth]{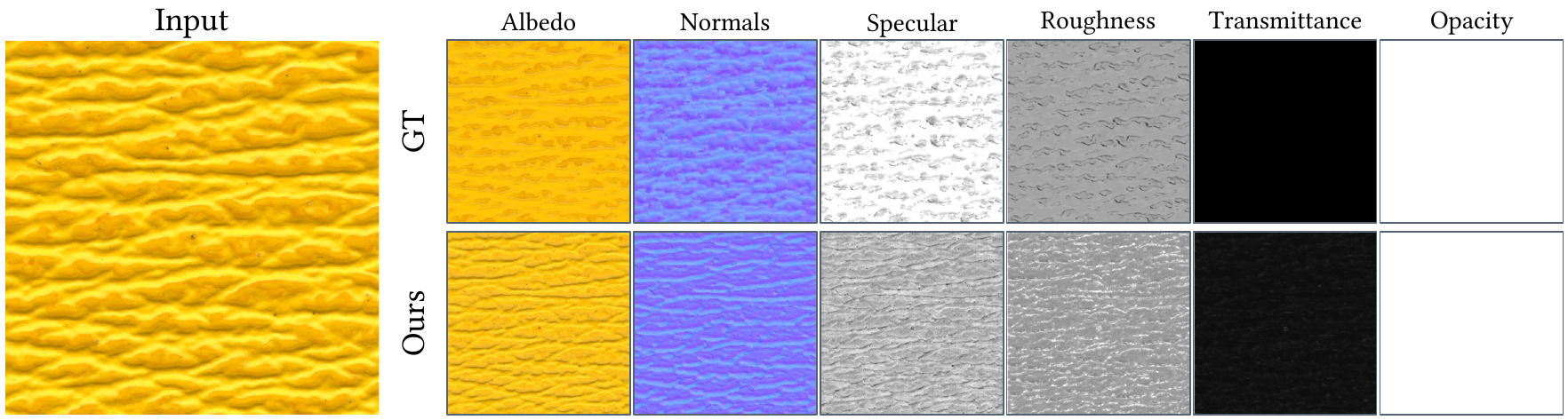}
	\caption{A failure case of our method. For the input image on the right, we show the ground truth albedo, normals, specular, roughness, transmittance and opacity maps (top row), and our model estimations. }
	\label{fig:failure}
\end{figure*}

Our method inherits the limitations of using flatbed scanners as a capture device. This setup cannot be used for non-flat materials (eg the marble in a statue) or materials which cannot physically be placed into this setup (eg a wall). It is also limited by our material model of choice, which, while it is more expressive than those of previous work, it cannot accurately represent complex phenomena such as anisotropy, strong displacements, high reflectivity, or subsurface scattering. Also, in order to capture non-uniform multi spectral absorption, $\mathbf{T}$ would require an additional attenuation value for each wavelength channel. Finally, our model sometimes struggles with some complex materials for which a single image is not a sufficient cue to estimate its optical properties. Such is the case for the bright, thick leather we show in Figure~\ref{fig:failure}, which is the material in our test set with the highest $\mathcal{L}_{\text{\tiny{BSDF}}}$.%
This type of material is also uncommon in our training dataset, which also explains the reduced generalization.

\bibliographystyle{cag-num-names}
\bibliography{references}

\end{document}